\documentclass{ws-procs9x6}
\begin{document}

\title{DRIFT EFFECTS AND THE AVERAGE FEATURES OF COSMIC RAY DENSITY GRADIENT IN CIRS DURING SUCCESSIVE TWO SOLAR MINIMUM PERIODS\\}
\author{A. FUSHISHITA$^{1}$, Y. OKAZAKI$^{2}$, T. NARUMI$^{1}$, C. KATO$^{1}$, S. YASUE$^{1}$, T. KUWABARA$^{3}$, J. W. BIEBER$^{3}$, P. EVENSON$^{3}$, M. R. DA SILVA$^{4}$, A. DAL LAGO$^{4}$, N. J. SCHUCH$^{5}$, M. TOKUMARU$^{6}$, M. L. DULDIG$^{7}$, J. E. HUMBLE$^{8}$, I. SABBAH$^{9}$, J. K\'{O}TA$^{10}$, and K. MUNAKATA$^{1}$}

\address{
$^{1}$Physics Department, Shinshu University, Matsumoto, Nagano 390-8621, Japan,\\
$^{2}$Department of Geophysics, Tohoku University, Sendai, Miyagi 980-0861, Japan\\
$^{3}$Bartol Research Institute and Department of Physics and Astronomy, University of Delaware, Newark, DE 19716, USA\\
$^{4}$National Institute for Space Research (INPE), 12227-010 Sao Jose dos Campos, SP, Brazil\\
$^{5}$Southern Regional Space Research Center (CRS/INPE), P.O. Box 5021, 97110-970, Santa Maria, RS, Brazil\\
$^{6}$Solar Terrestrial Environment Laboratory, Nagoya University, Nagoya, Aichi 464-8601, Japan\\
$^{7}$Australian Antarctic Division, Kingston, Tasmania 7050, Australia\\
$^{8}$School of Mathematics and Physics, University of Tasmania, Hobart, Tasmania 7001, Australia\\
$^{9}$Department of Physics, Faculty of Science, Kuwait University, Kuwait; on leave from Physics Department, Faculty of Science, University of Alexandria, Egypt\\
$^{10}$Lunar and Planetary Laboratory, University of Arizona, Tucson, AZ 87721, USA}
%############### Abstract ################
\begin{abstract}
We deduce on hourly basis the spatial gradient of the cosmic ray density in three dimensions from the directional anisotropy of high-energy ($\sim$50 GeV) galactic cosmic ray (GCR) intensity observed with a global network of muon detectors on the Earth's surface. By analyzing the average features of the gradient in the corotational interaction regions (CIRs) recorded in successive two solar activity minimum periods, we find that the observed latitudinal gradient ($G_z$) changes its sign from negative to positive on the Earth's heliospheric current sheet (HCS) crossing from the northern to the southern hemisphere in A$<$0 epoch, while it changes from positive to negative in A$>$0 epoch. This is in accordance with the drift prediction. We also find a negative enhancement in $G_x$ after the HCS crossing in both A$<$0 and A$>$0 epochs, but not in $G_y$. This asymmetrical feature of $G_x$ and $G_y$ indicates significant contributions from the parallel and perpendicular diffusions to the the gradient in CIRs in addition to the contribution from the drift effect. 
\end{abstract}

\bodymatter
%############### Introduction ################
\section{Introduction}

The drift model of cosmic ray transport in the heliosphere predicts a bidirectional latitude gradient of the GCR density, pointing in opposite directions on opposite sides of the HCS [\cite{Jokipii82}][\cite{Kota83}]. The predicted spatial distribution of the GCR density has a minimum along the HCS in the ``positive'' polarity period of the solar polar magnetic field (also referred as the A$>$0 epoch), when the interplanetary magnetic field (IMF) directs away from (toward) the Sun in the northern (southern) hemisphere, while the distribution has the local maximum on the HCS in the ``negative'' period (A$<$0) with the opposite field orientation in each hemisphere. The field orientation reverses every 11 years around the maximum period of solar activity. By analyzing the GCR density measured with the anticoincidence guard on board {\it ACE} and {\it Helios} satellites and by ground-based neutron monitors (NMs), Richardson {\it et al}. [\cite{Richardson96}] and Richardson [\cite{Richardson04}] investigated GCR modulation by CIRs, which are formed at the leading edges of corotating high-speed solar wind streams originating in coronal holes on the solar surface. From statistical analyses of variations in many CIRs with and without the HCS, the above authors concluded that the IMF sector boundary (i.e., HCS) does not organize the GCR density, contrary to expectations from the simple drift model. On the other hand, they also found observational evidence that the amplitude of the CIR-related modulations of the GCR density around the solar minimum is $\sim$ 50\% larger in A$>$0 epochs than in A$<$0 epochs. The GCR modulation in CIRs is still an open question.

These analyses were based on a one dimensional spatial distribution of the GCR density measured as the temporal variation of the GCR count rate recorded by each detector as it sweeps across the CIR. A single detector measurement of the density cannot, however, observe the spatial distribution separately from the temporal variation of the density at a particular location in space. The spatial distribution sampled by a single detector also does not reflect the global distribution surrounding the detector, above and below the detector's viewing path. It is also affected by local irregularities in the solar wind. This can be serious, particularly when analyzing satellite measurements of low-energy (sub-GeV) GCRs.

The spatial distribution of GCRs also can be inferred from the GCR spatial gradient in three dimensions, which provides us with additional information on the distribution around the detectors¢ viewing paths. The diffusive flux equation describes the cosmic ray anisotropy in terms of the solar wind convection and the GCR density gradient. By inversely solving the diffusive flux equation, including the drift, Okazaki {\it et al.} [\cite{Okazaki08}] (hereafter referred to as paper I) deduced the gradient from the anisotropy that is derived from the observation made by the Global Muon Detector Network (GMDN). The GMDN has a major response to 50GeV GCRs whose Larmor radii are as large as 0.2AU in 5nT IMF. They analyzed the three geocentric solar eclipic coordinate (GSE) components of the gradient observed in association with CIRs in a single solar rotation period (Carrington Rotation CR 2043) in 2006, around the solar activity minimum of A$<$0 epoch. It was found that the temporal evolution of the gradient before and after the HCS crossing is consistent with the prediction of the drift model, that the GCR distribution has a local maximum on the HCS [\cite{Jokipii82}][\cite{Kota83}]. As far as the gradient in this particular rotation period concerned, on the other hand, there is no clear signature of the significant modulation in association with enhancements in the solar wind velocity and the IMF magnitude in the CIR. Examining the drift effect solely from the gradient  in a single rotation in A$<$0 epoch, however, is not easy, as the local maximum of the GCR density on the HCS is expected not only from the drift effect, but also from the minimum solar wind velocity on the HCS in a CIR. In the present paper, therefore, we examine the drift effect by analyzing the average gradient of multiple CIRs observed during the minimum periods of the solar activity in A$<$0 and A$>$0 epochs. 

\begin{figure}
\begin{center}
\psfig{file=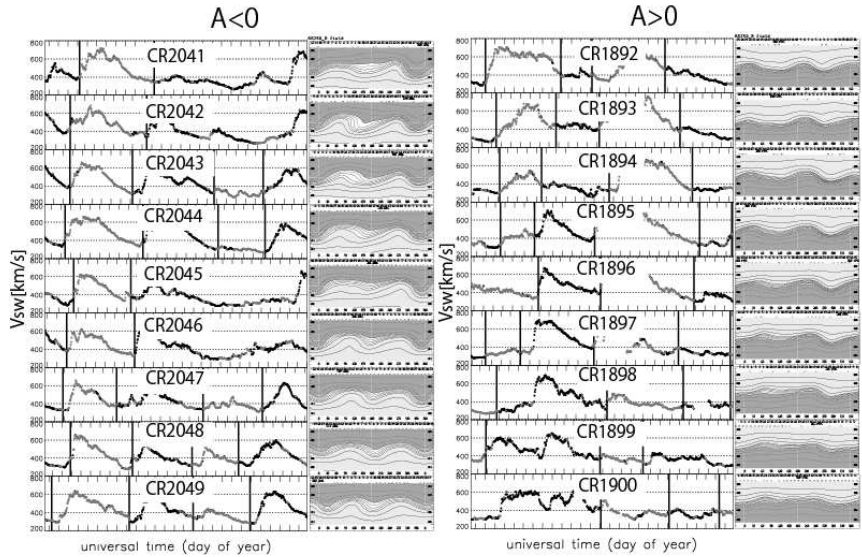,width=4.5in}
\end{center}
\caption{Sample HCSs superposed in this paper for A$<$0 (left) and A$>$0 (right) epochs. Each panel displays the solar wind velocity in a solar rotation period as a function of the universal time (day) on the horizontal axis. The time of the Earth's HCS crossing is indicated by the vertical solid line, while the away and toward IMF sectors are indicated by black and gray data points, respectively. Attached on the right side in each figure is the source surface synoptic charts (Radial 250) in the corresponding rotation by the Wilcox Solar Observatory$^c$}
\label{fig1}
\end{figure}

\section{Analysis and result}
\subsection{Superposition analysis of CIRs at the HCS crossing}
\begin{figure}
\begin{center}
\psfig{file=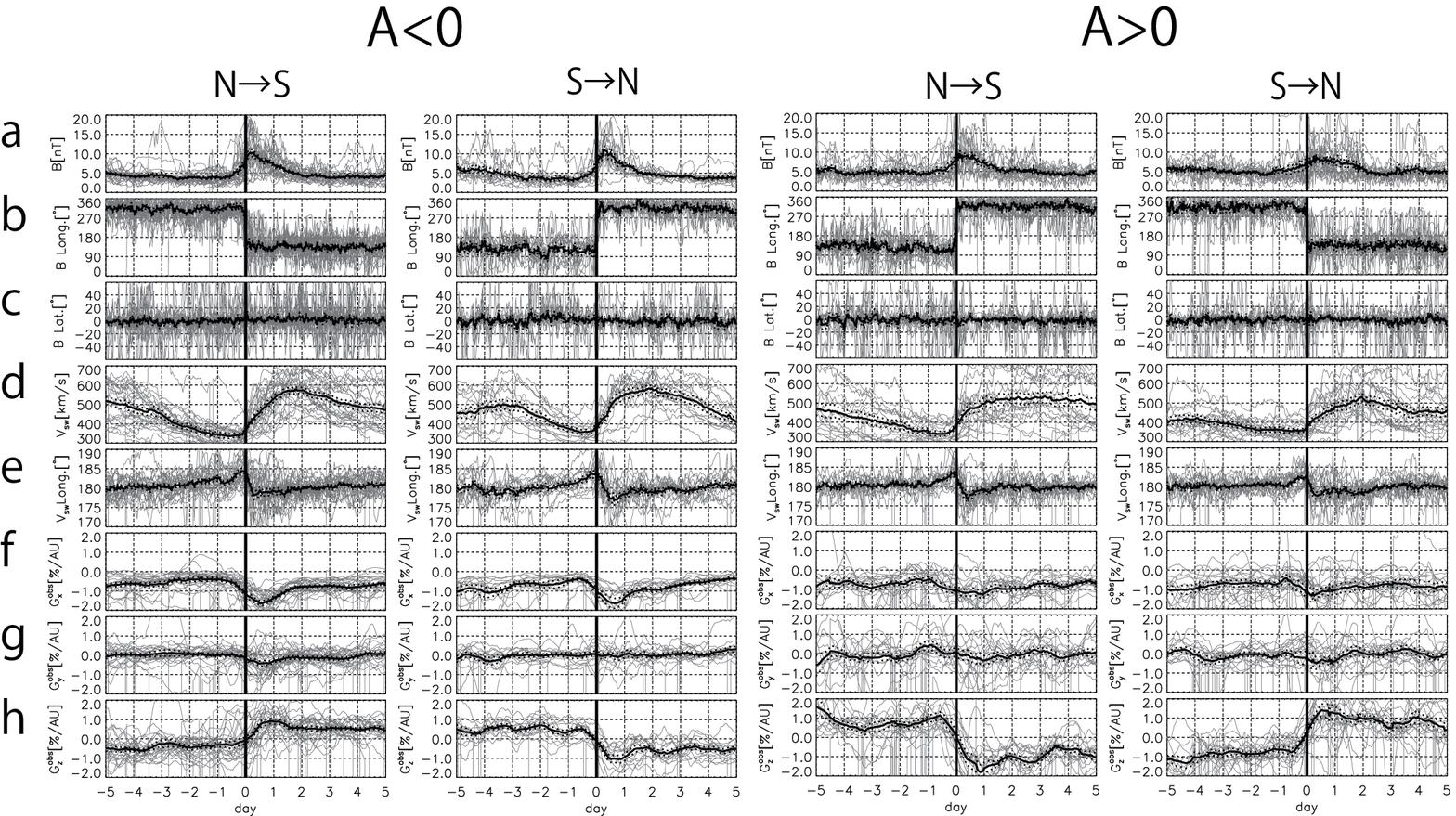,width=4.5in}
\end{center}
\caption{From top to bottom, each panel shows the IMF magnitude (a), the GSE-longitude and latitude of the IMF (b-c), the solar wind velocity (d), the GSE-longitude of the solar wind (e) and three GSE components of the observed gradient (f-h). CIRs superposed at the HCS crossing in A$<$0 (left) and A$>$0 (right) epochs. Thin gray curves in each panel display individual CIRs superposed, while the black curve shows the average as a function of time from the HCS crossing indicated by a vertical solid line. The dotted curves above and below the black curve bound the error of the average evaluated from the standard deviation of superposed CIRs. Only $\pm$5 days period from the HCS crossing is plotted. The superposition at the HCS crossing from the northern (southern) hemisphere to the southern (northern) hemisphere is indicated by ''N$\rightarrow$S'' (''S$\rightarrow$N'') at the top of each figure. 
}
\label{fig2}
\end{figure}

We obtain the average feature of the GCR density gradient by superposing multiple CIRs at the timing of the HCS crossing in both A$<$0 and A$>$0 epochs. Throughout this paper, we use the hourly mean solar wind velocity and IMF data from the Omni2 data for A$>$0 epoch and from {\it ACE}-Level2 data for A$<$0 epoch, respectively\footnote{Omni2 data are available at http://omniweb.gsfc.nasa.gov/html/ow\_data.html}\footnote{{\it ACE}-Level2 data are available at http://www.srl.caltech.edu/ACE/ASC/}. We use {\it ACE}-Level2 data lagged by one hour, as a rough correction for the solar wind transit time between {\it ACE} and the Earth. By using IMF components ($B_x$,$B_y$,$B_z$) in GSE coordinates, the IMF sector is designated {\it toward} ({\it T}) if $B_x>B_y$ and {\it away} ({\it A}) if $B_y>B_x$. We define the HCS crossing at the timing when the sector polarity ({\it toward} or {\it away}) designated by 23-hours central moving average of $B_x$ and $B_y$ alternates. We then choose HCS crossings associated by the enhancement of the solar wind velocity with the maximum velocity exceeding 500 km/s and the IMF magnitude with the maximum exceeding 10 nT. According to this selection criteria, we chose 42 CIRs observed in 2006-2008 for A$<$0 period and 38 CIRs in 1995-1998 for A$>$0 epoch, respectively. The number of CIRs superposed for A$>$0 period is limited mainly due to many data gaps in the Omni2 data. Figure \ref{fig1} displays the solar wind velocity and the source surface synoptic charts (Radial 250 model) by the Wilcox Solar Observatory\footnote{The source surface synoptic charts and the HCS tilt angle are available at http://wso.stanford.edu/}. The HCS crossings identified in this paper are also indicated by the vertical solid line in this figure. The average HCS tilt angle in A$<$0 epoch analyzed in this paper is 15.2$^{\circ}$, while it is 9.0$^{\circ}$ in A$>$0 epoch. The superposed IMF and solar wind velocity are displayed in Figures \ref{fig2}a-e as functions of time for $\pm$5 days from the HCS crossing. Clear signatures of CIR around the timing of the HCS crossing, such as enhancements of the IMF magnitude and the solar wind velocity (Figures \ref{fig2}a and \ref{fig2}d) and reversals of the azimuthal solar wind flow angle (Figure \ref{fig2}e), are clearly seen in this figure. We also note that the superposed IMF longitude in {\it away} ({\it toward}) sector in Figure \ref{fig2}b is close to 135$^{\circ}$ (315$^{\circ}$) and the average latitude in Figure \ref{fig2}c is close to 0$^{\circ}$ regardless the IMF sector. This implies that the nominal Parker spiral field with the spiral angle of 45$^{\circ}$ is a good approximation of the observed IMF as long as the average field is concerned.

\subsection{Analysis of GCR density gradient in three dimensions}\label{analysisGCR}

We deduce the GCR density gradient in three dimensions on hourly basis from the anisotropy observed with the network of muon detectors. Readers can refer to paper I for each muon detector in the network and also for the detail analysis method to derive the gradient from the observed anisotropy. In this paper, we analyze the average gradient in multiple CIRs observed in 2006-2008 with the GMDN during the minimum period of the solar activity in A$<$0 epoch. We also derive the average gradient in 1995-1998 in A$>$0 epoch by using the observation with a couple of muon detectors at Nagoya and Hobart and examine the dependence of the gradient on the polarity of the solar magnetic field.

We first correct the observed anisotropy vector in GSE coordinates for the solar wind convection anisotropy and the Compton-Getting anisotropy arising from the Earth's 30 km/s orbital motion. We then divide the corrected anisotropy vector {\mbox{\boldmath{$\xi$}}}$(t)$ into components parallel and perpendicular to the IMF as
\begin{equation}
{\mbox{\boldmath{$\xi$}}}(t)={\mbox{\boldmath{$\xi_\parallel$}}}(t)+{\mbox{\boldmath{$\xi_\perp$}}}(t).
\end{equation}
These components are given as
\begin{equation}
{\mbox{\boldmath{$\xi_\parallel$}}}(t)=R_{L}(t)\alpha_{\parallel}{\bf{G}}_{\parallel}^{Diff}(t), \label{Analysis03}
\end{equation}
\begin{equation}
{\mbox{\boldmath{$\xi_{\perp}$}}}(t)=R_{L}(t)[\alpha_{\perp}{\bf{G}}_{\perp}(t)-{\bf{b}}(t)\times{\bf{G}}_{\perp}(t)],  \label{Analysis04}
\end{equation}
where ${\bf{G}}_{\parallel}^{Diff}(t)$ and ${\bf{G}}_{\perp}(t)$ are respectively the density gradient components parallel and perpendicular to the IMF, $R_L(t)$ is the particle's effective Larmor radius of 50GeV protons in the observed IMF which is 0.2 AU in 5nT IMF, ${\bf{b}}(t)$ is the unit vector pointing in the direction of the IMF, and $\alpha_{\parallel}$ and $\alpha_{\perp}$ are the dimensionless mean free paths ($\lambda_{\parallel}(t)$ and $\lambda_{\perp}(t)$) of GCR scattering by magnetic irregularities, defined as
\begin{equation}
\alpha_{\parallel}=\lambda_{\parallel}/R_L (t),\label{Analysis05}
\end{equation}
\begin{equation}
\alpha_{\perp}=\lambda_{\perp}/R_L (t).\label{Analysis06}
\end{equation}

In this paper, we adopt an ad hoc choice of $\alpha_{\parallel}=7.2$ and $\alpha_{\perp}=0.36$ which were also assumed in paper I. Qualitative features would remain similar for a wide range of reasonable parameters. Inversely solving Eqs.(\ref{Analysis03}) and (\ref{Analysis04}) for ${\bf{G}}_{\parallel}^{Diff}(t)$ and ${\bf{G}}_{\perp}(t)$, we obtain
\begin{equation}
{\bf{G}}_{\parallel}^{Diff}(t)=\frac{1}{\alpha_{\parallel}R_{L}(t)}{\mbox{\boldmath{$\xi_\parallel$}}}(t),\label{Analysis07}
\end{equation}
\begin{equation}
{\bf{G}}_{\perp}(t)={\bf{G}}_{\perp}^{Diff}(t)+{\bf{G}}_{\perp}^{Drift}(t) \label{Analysis09}
\end{equation}
where,
\begin{equation}
{\bf{G}}_{\perp}^{Diff}(t)=\frac{\alpha_{\perp}}{(1+\alpha_{\perp}^2)R_{L}(t)} {\mbox{\boldmath{$\xi_\perp$}}}(t) \label{Analysis10}
\end{equation}
and,
\begin{equation}
%{\bf{G}}_{\perp}^{Drift}(t)=\frac{1}{(1+\alpha_{\perp}^2)R_{L}(t)}{\bf{b}}(t)\times{\mbox{\boldmath{$\xi_{\perp}$}}}(t). \label{Analysis11}
{\bf{G}}_{\perp}^{Drift}(t)=\frac{1}{(1+\alpha_{\perp}^2)R_{L}(t)}{\bf{b}}\times{\mbox{\boldmath{$\xi_{\perp}$}}}(t). \label{Analysis11}
\end{equation}
Finally ${\bf{G}}(t)$ can be expressed in terms of two distinct contributions respectively from the diffusion and drift effects as,
\begin{equation}
{\bf{G}}(t)=\{{\bf{G}}_{\parallel}^{Diff}(t)+{\bf{G}}_{\perp}^{Diff}(t)\}+{\bf{G}}_{\perp}^{Drift}(t), \label{Analysis12}
\end{equation}
where the first two terms in the bracket on the right hand side express the contribution from the parallel and perpendicular diffusions respectively, while the second term denotes the contribution from the drift effect.

\begin{figure}
\begin{center}
\psfig{file=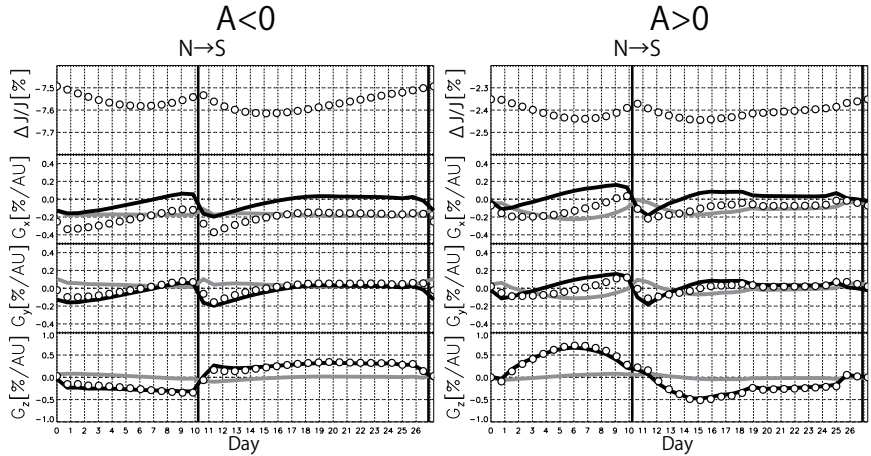,width=4.5in}
\end{center}
\caption{The numerical solution of a simple drift model for 50 GeV GCRs. From top to bottom, each panel displays the GCR density and three GSE components of the gradient as a function of day in a solar rotation period in A$<$0 (left) and A$>$0 (right) epochs. The numerical values are shown by open circles. Contribution from the diffusion effect is shown by gray curve, while the contribution from the drift effect is displayed by black curve. The HCS crossing is indicated by a vertical black line. See paper I for the detail of this numerical model.}
\label{fig3}
\end{figure} 

Three GSE-components of the superposed ${\bf{G}}(t)$ are displayed in Figures \ref{fig2}f-h, while the numerical prediction of a simple drift model are shown in Figure \ref{fig3}. This model assumes that the solar wind velocity and IMF magnitude are constant without any enhancement seen as CIR signatures. This model is identical to that which paper I compared with the observation and readers can refer to paper I for more detail of this model. The observed gradient ${\bf{G}}(t)$ in Eq. (\ref{Analysis12}) shows a large fluctuation due to the large fluctuation of the IMF orientation ($b_x$, $b_y$, $b_z$). To show the systematic variation of each component, we filter this fluctuation with a central 23 hour moving average of hourly data. In Figure \ref{fig2}, the gradient in each CIR event is displayed by a gray curve, while the average gradient of all CIR events is plotted as the black curve. It is clear that the average GSE-z component of the gradient ($G_z$) in Figure \ref{fig2}h is negative (positive) above (below) the HCS in A$<$0 epoch, while it is positive (negative) above (below) the HCS in A$>$0 epoch. This is consistent with the drift model prediction in Figure \ref{fig3} at least qualitatively. We confirm this for the first time with the average gradient in multiple CIRs observed. It is also seen that the GSE-x component of the average gradient ($G_x$) remains negative in Figures \ref{fig2}f implying that the ambient radial density gradient is positive and the density increases away from the Sun in response to the outward convection by the solar wind.
\begin{figure}
\begin{center}
\psfig{file=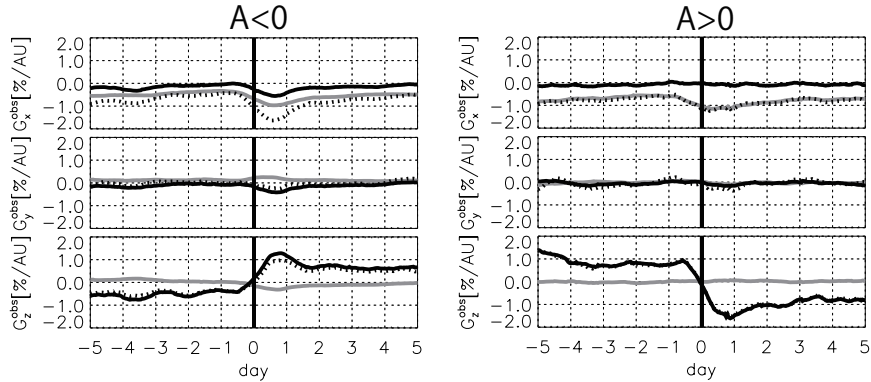,width=4.5in}
\end{center}
\caption{Contributions from the diffusion and drift effects to three GSE components of the average gradient ${\bf{G}}(t)$. From top to bottom, dotted curve in each panel displays the average $G_x$, $G_y$ and $G_z$ in Figure \ref{fig2}, while gray and black curves show respectively the contributions to ${\bf{G}}(t)$ from the diffusion (${\bf{G}}_{\parallel}^{Diff}(t)+{\bf{G}}_{\perp}^{Diff}(t)$) and drift (${\bf{G}}_{\perp}^{Drift}(t)$) effects, in A$<$0 epoch (left) and A$>$0 epoch (right). $G_x$ and $G_y$ in this figure are obtained by averaging those in ''N$\rightarrow$S'' and ''S$\rightarrow$N'' in Figures \ref{fig2}f and \ref{fig2}g. We also calculate the average for $G_z$ in Figure \ref{fig2}h after reversing sign of $G_z$ in ''S$\rightarrow$N'' in Figure \ref{fig2}h.}
\label{fig4}
\end{figure}

In Figure \ref{fig2}f, we also note the negative enhancement in $G_x$ observed following the HCS crossing. Such enhancements were first observed in both $G_x$ and $G_y$ in a sample rotation period in paper I and attributed to the drift effect causing the bidirectional density gradient toward the HCS in the ecliptic plane, together with the contribution from the ambient radial gradient away from the Sun. Similar enhancements are also seen in both  $G_x$ and $G_y$ in the drift model prediction in Figure \ref{fig3}. The observed gradient in Figure \ref{fig2}, however, shows the enhancement only in $G_x$, but not in $G_y$. This asymmetry is interpreted in terms of the contribution from the diffusion effect as follows. Figure \ref{fig4} shows three components of the average ${\bf{G}}(t)$ by dotted curves, together with contributions from the diffusion effect (${\bf{G}}_{\parallel}^{Diff}(t)+{\bf{G}}_{\perp}^{Diff}(t)$) by gray curves and the drift effect (${\bf{G}}_{\perp}^{Drift}(t)$) by black curves in Eq. (\ref{Analysis12}). In A$<$0 epoch, ${\bf{G}}_{\parallel}^{Diff}(t)$ due to the inward diffusion along the IMF is dominant as well as ${\bf{G}}_{\perp}^{Drift}(t)$. ${\bf{G}}_{\perp}^{Drift}(t)$ shows the negative enhancement in both $G_x$ and $G_y$, while ${\bf{G}}_{\parallel}^{Diff}(t)$ shows the negative enhancement in $G_x$, but the positive enhancement in $G_y$. This is because of the geometry of the nominal Parker field in the GSE coordinate system. This asymmetric contribution from ${\bf{G}}_{\parallel}^{Diff}(t)$ to $G_x$ and $G_y$ results in the asymmetry seen in Figure \ref{fig2}. This asymmetry, on the other hand, is less evident in A$>$0 epoch, because of another contribution from ${\bf{G}}_{\perp}^{Diff}(t)$ due to the inward perpendicular diffusion which results in negative enhancements in both $G_x$ and $G_y$. These contributions to the observed ${\bf{G}}(t)$ from the parallel and perpendicular diffusions are not clarified by a sample rotation analysis in paper I and first revealed in this paper by analyzing the average gradient in multiple CIR events. We are now preparing for a quantitative analyses of each contribution from ${\bf{G}}_{\parallel}^{Diff}(t)$, ${\bf{G}}_{\perp}^{Diff}(t)$ and ${\bf{G}}_{\perp}^{Drift}(t)$.
\begin{figure}
\begin{center}
\psfig{file=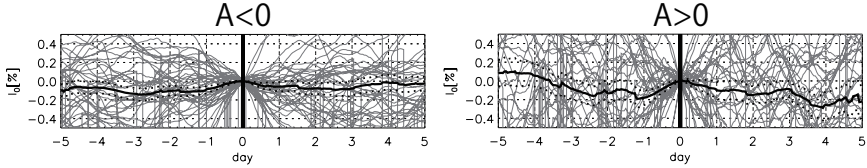,width=4.5in}
\end{center}
\caption{GCR density superposed at the HCS crossing in A$<$0 (left) and A$>$0 (right) epochs. This figure is plotted in the same format as Figure \ref{fig2}, except that the derrived density is normalized to 0 \% at the HCS crossing to clarify the variation associated with the HCS crossing.}
\label{fig5}
\end{figure}

In Figure \ref{fig5}, we finally display the GCR density which is derived as the omnidirectional intensity, together with the anisotropy vector, from the best-fitting to the observed muon count rates in various directional channels (see paper I). In this figure, we normalize the density to 0 \% at the HCS crossing to clarify the variation associated with the HCS crossing. It is seen that the amplitude of the average variation is small and the fluctuation is large even for 50 GeV GCRs recorded with muon detectors, easily masking the average systematic variation seen in figure \ref{fig3} by the numerical model. This difficulty due to the fluctuation can be serious, particularly when analyzing low-energy GCRs, which have smaller Larmor radii and are more sensitive to small-scale structures in the solar wind. The variation in this figure is much less significant than the average gradient variation in Figure \ref{fig2}. This implies how difficult is to deduce the GCR spatial distribution solely from single detector measurement of the GCR density[\cite{Richardson96}][\cite{Richardson04}].

%############## discussions
\section{Discussions}

In table 1, we compare the average GCR gradient and density with the model prediction. We first note that the magnitude of $G_z$ is 50 \% larger in A$>$0 epoch in accordance with the model prediction. The magnitude of negative $G_x$, on the other hand, is larger in A$>$0 epoch, while the model predicts $G_x$ almost twice in A$<$0 epoch in an oppsite sense. We also note the average density ($I_0$) being 50 \% lower in A$>$0 epoch. As we normalized $I_0$ to 0 \% at the HCS crossing, this implies the amplitude of the density variation before and after the HCS crossing is 50 \% larger in A$>$0 epoch, in accordance with Richardson {\it et al.} [\cite{Richardson04}] reporting that the amplitude of the recurrent GCR intensity variations in solar rotation periods are 50 \% larger in A$>$0 epoch than in A$<$0 epoch.

In concluding anything from these observed features about the drift model prediction, however, we should be aware of that the gradient in A$>$0 epoch presented in this paper might be biased, because it is derived from an incomplete network before the GMDN started operation in March, 2006. To check such a possible bias, we analyzed A$<$0 epoch by using only two detectors (at Nagoya and Hobart) used for A$>$0 in the GMDN and confirmed that the derived gradient can be $10\sim30$ \% larger than that derived from the entire GMDN depending on values of $\alpha_{\parallel}$ and $\alpha_{\perp}$ assumed. In order to precisely clarify the difference of the gradient in A$<$0 and A$>$0 epochs, therefore, we need to improve the analysis method further and correct the observed gradient in A$>$0 for this bias effect.

%=== Table 1 ===
\begin{table}
\begin{center}
\tbl{GCR density gradient ($G_x$,$G_y$,$G_z$), density ($I_0$), solar wind velocity ($V_{SW}$) and IMF magnitude (B) averaged over a period in Figure \ref{fig2} in A$<$0 and A$>$0 epochs. The upper lines for the gradient and density represent the observed average, while the lower lines represent the model prediction. The average $G_z$ is calculated after reversing the sign of $G_z$ in {\it toward} sectors. The period between $\pm$1 day from the HCS crossing is excluded from the average calculation. Errors for the observed values are deduced from the standard deviation of CIRs used for the calculation, while errors for the model prediction are derived from the standard deviation of data points in Figure \ref{fig3}.}
{
\begin{tabular}{ccccccccccc}
\hline
& &Heliomagnetic polarity & & A$<$0 & & A$>$0 & & & &\\
\hline \hline
% component    A<0                A<>0
& & $G_x[\%/AU]$ & & {\bf -0.706$\pm$0.006} & & {\bf -0.781$\pm$0.009} & & & &\\
\cline{4-8}
& &              & & -0.215$\pm$0.016 & & -0.105$\pm$0.017 & & & &\\
\hline
& & $G_y[\%/AU]$ & & {\bf -0.017$\pm$0.005} & & {\bf -0.060$\pm$0.009} & & & &\\
\cline{4-8}
& &              & & -0.005$\pm$0.015 & & -0.003$\pm$0.016 & & & &\\
\hline
& & $G_z[\%/AU]$ & & {\bf  0.585$\pm$0.005} & & {\bf  0.890$\pm$0.009} & & & &\\
\cline{4-8}
& &              & &  0.239$\pm$0.017 & &  0.358$\pm$0.047 & & & &\\
\hline
& & $I_0[\%]$    & & {\bf -0.077$\pm$0.004} & & {\bf -0.112$\pm$0.008} & & & &\\
\cline{4-8}
& &              & & -0.042$\pm$0.005 & & -0.039$\pm$0.006 & & & &\\
\hline
& & $V_{SW}[km/s]$ &  & {\bf 482.74$\pm$1.18} & & {\bf 449.01$\pm$1.32} & & & &\\
\hline
& & B[nT]        & & {\bf 4.373$\pm$0.021} & &  {\bf 5.112$\pm$0.029} & & & &\\
\hline
\end{tabular}
}
\end{center}
\label{tab1}
\end{table}

\section*{Acknowledgments}

This work is supported in part by U.S. NSF grant ATM-0527878 and NASA grant NNX07-AH73G, and in part by Grants-in-Aid for Scientific Research from the Ministry of Education, Culture, Sports, Science and Technology in Japan and by the joint research program of the Solar-Terrestrial Environment Laboratory, Nagoya University. The observations with the Kuwait University muon detector are supported by the Kuwait University grant SP03/03. We thank N. F. Ness for providing {\it ACE} magnetic field data via the {\it ACE} Science Center.

\end{document}